\documentclass [aps,prb,10pt,twocolumn,showpacs,superscriptaddress]{revtex4-1}
\usepackage[english]{babel}
\usepackage{graphicx}
\usepackage{amsmath,amssymb}
\usepackage[utf8]{inputenc}
\usepackage{natbib}
\usepackage{units}
\usepackage{hyperref}
\usepackage{color}
\usepackage{url}
\hypersetup{colorlinks=true,citecolor={blue},linkcolor={blue},urlcolor={blue}}
\usepackage{balance}
\usepackage{textcomp}
\usepackage{epstopdf}

\makeindex

\begin{document}
\title{Parity bifurcations in trapped multistable phase locked exciton-polariton condensates}

\date{\today}

\author{E. Z. Tan}
\email[correspondence address:~]{TANE0034@e.ntu.edu.sg}
\affiliation{Division of Physics and Applied Physics, School of Physical and Mathematical Sciences, Nanyang Technological University 637371, Singapore}

\author{H. Sigurdsson}
\affiliation{Science Institute, University of Iceland, Dunhagi-3, IS-107 Reykjavik, Iceland}
\author{T. C. H. Liew}
\affiliation{Division of Physics and Applied Physics, School of Physical and Mathematical Sciences, Nanyang Technological University 637371, Singapore}

\begin{abstract}
We present a theoretical scheme for multistability in planar microcavity exciton-polariton condensates under nonresonant driving. Using an excitation profile resulting in a spatially patterned condensate, we observe organized phase locking which can abruptly reorganize as a result of pump induced instability made possible by nonlinear interactions. For $\pi/2$ symmetric systems this reorganization can be regarded as a parity transition and is found to be a fingerprint of multistable regimes existing over a finite range of excitation strengths. The natural degeneracy of the planar equations of motion gives rise to parity bifurcation points where the condensate, as a function of excitation intensity, bifurcates into one of two anisotropic degenerate solutions. Deterministic transitions between multistable states are made possible using controlled nonresonant pulses, perturbing the solution from one attractor to another.
\end{abstract}

\maketitle
\section{Introduction}

A microcavity exciton-polariton is a bosonic quasiparticle that arises from the strong coupling between a quantum well exciton and a cavity photon \cite{yamamoto2002}. Due to their bosonic statistics and short lifetimes, polaritons can form a non-equilibrium analog of a Bose-Einstein condensate at high nonresonant excitation intensities where stimulated scattering results in large coherent population balanced by gain and decay. Their ability to interact strongly with themselves gives rise to a $\chi^{(3)}$ Kerr-like nonlinearity which in turn, for resonant excitation schemes on scalar condensates, gives rise to optical bistability \cite{baas2004optical, gippius2004nonlinear, Baas_PRB2004, whittaker2005effects}, spatial multistability \cite{Plamondon_PRB2016}, and multistability for vectorial (spin dependent) condensates \cite{paraiso2010multistability, gippius2007polarization,gavrilov2010multistability}.

Recently, ultrafast~\cite{cerna2013ultrafast, de2012control}, ultralow energy~\cite{Dreismann_NatMat2016}, and nonlinear relaxation~\cite{grosso2014nonlinear} switching mechanisms were realized experimentally between bistable states of polariton ensembles and suggest their future application as optical memory elements. Bistability is also a precursor for a number of effects in exciton-polariton systems, including: the formation of solitons \cite{larionova2008optical, egorov2010two, egorov2013formation}; imprinting patterns and spatial images \cite{sarkar2010polarization, adrados2010spin, sekretenko2013spin, amo2011polariton}; screening disorder \cite{liew2012optically}; sustaining superfluid propagation \cite{pigeon2017sustained}; realizing quasi-compactons \cite{kartashov2012compactons}; simulating cellular automata \cite{li2016cellular}; realizing emergent descriptions of (classical) Ising models \cite{foss2017emergent}; and realizing polaritonic circuits \cite{espinosa2013complete}.

While optical multistability with resonant excitation schemes have been the subject of intensive research \cite{baas2004optical, gippius2004nonlinear, Baas_PRB2004, whittaker2005effects,paraiso2010multistability, gippius2007polarization,gavrilov2010multistability,Plamondon_PRB2016}, optical multistability in nonresonantly excited microcavity polaritons is less well studied. Physically, the nonresonant pump scheme is advantageous since a resonant pump scheme is implemented with an external laser, but nonresonant pumping can be achieved with electrical contacts on both sides of the quantum well cavity to spur the creation of excitons which is far more compact to implement in optoelectronic circuits. To date, theoretical schemes for incoherently excited bistability based on strong saturated absorption \cite{karpov2015dissipative}, thermally-induced changes \cite{cotta2007bistability}, modulational instability \cite{kyriienko2014bistability}, and gain competition between symmetric and antisymmetric condensate modes \cite{sigurdsson2015switching} have been proposed. Fairly recently, nonresonant optical bistability was demonstrated in polariton condensates under electrical injection of charge carriers \cite{Amthor_PRB2015, Klaas_PRB2017} and between spin polarizations in annular pumping geometries \cite{Berloff_SpinBist2017}.

In this paper, we investigate the appearance of optical multistability in a planar two-dimensional (2D) exciton-polariton condensate under spatially patterned nonresonant pumping. In Ref.~[\onlinecite{sigurdsson2015switching}] optical bistability has been shown theoretically to arise due to the modulational instability between the parities of a localized one dimensional (1D) polariton condensate pattern; each bistable state corresponding either to an antisymmetric or symmetric macroscopic wavefunction. Consequently, it stands to reason that multistability can be observed in higher dimensional geometries due to the increased degeneracy of the linear part of the equations of motion. 

The multistability relies on pump induced parity cross-saturation previously considered in 1D exciton-polariton condensates~\cite{sigurdsson2015switching, Sigurdsson_PRB2017}. The symmetry of the dynamical equations results in a condensation threshold belonging to a definite parity of the system due to optimal constructive interference between same parity modes (phase locked condensates). A result well studied in the field of nonlinear optics~\cite{Wang_APL1988} and in agreement with recent experimental observations where different excitation geometries result in exciton-polariton condensates being phase locked either in-phase or anti-phase~\cite{Tosi_NatComm2012, Cristofolini_PRL2013, Ohadi_PRX2016, Lagoudakis_NJP2017, Kalinin_ArXiv2017_2, Kalinin_ArXiv2017}. Increasing the pump intensity beyond threshold results in an unstable condensate which rapidly transitions into a new solution of different parity, corresponding to a reorganization of phase lockings within the system. Sweeping backwards in pump intensity, several hystereses are recovered between the condensate parities. Within these hystereses we find that regimes of multistability between anisotropic degenerate solutions exist.

We classify a point of instability where an isotropic condensate transitions into one of two anisotropic degenerate condensate solutions with equal probability, a feature which can be regarded as \emph{parity bifurcation}. Very recently, bifurcating points were observed in spinor polariton condensates where, also as a function of pump intensity, a linearly polarized condensate transitions to either of two circularly polarized states~\cite{Ohadi_2015PRX}. 

After characterizing the possible multistable regimes and states therein, we proceed to demonstrate a method to induce controlled, reversible transitions within a group of stable states based on the application of nonresonant excitation pulses.

\section{\label{Model} Condensate phase locking}
Eigensolutions of spatially symmetric Hamiltonians are classified as either symmetric (even) or antisymmetric (odd). For nonlinear equations of motion this is no longer necessarily the case. Regardless, phase locking between spatially separate condensates~\cite{Tosi_NatComm2012, Cristofolini_PRL2013, Ohadi_PRX2016} still takes place, made possible through their interactions, where condensates form either symmetric or antisymmetric states due to constructive interference of polaritons traveling from one condensate to another. Due to the short polariton lifetime, there naturally exists only a finite range of distances between the two condensates to coherently phase lock before traveling polaritons decay away.

We consider a non-equilibrium condensate of 2D planar exciton-polaritons described by the driven-dissipative nonlinear Schrödinger equation. Standard heuristics dictate that the feeding of polaritons into the condensate is controlled by active-exciton reservoir rate equations \cite{Wouters_PRL2007}. Under the assumption of a fast active-exciton reservoir dephasing rate and low polariton densities the equation of motion for lower polaritons in the parabolic regime can be simplified into the following \cite{Keeling_PRL2008}:
\begin{align} \notag
i\hbar\frac{\partial \psi }{\partial t} = & \bigg[ -\frac{\hbar^{2} \nabla^2}{2m} +  V\left( \mathbf{r} \right)  + iP\left( \mathbf{r} \right)  \\ \label{SE}
 - & i \frac{\hbar \Gamma}{2} + \left( \alpha -iR \right) \left| \psi \right|^{2} \bigg]\psi .
\end{align}
Here $\psi$ is the polariton condensate macroscopic wavefunction or order parameter; $\nabla^2$ is the 2D Laplacian; $m$ is the polariton effective mass; $P(\mathbf{r})$ is a spatially dependent nonresonant feeding of polaritons into the condensate; $V(\mathbf{r}) = V_0(\mathbf{r}) + g P(\mathbf{r})$ is the potential landscape of the system corresponding to an external potential and pump induced blueshift respectively; $\Gamma$ is the average dissipation rate; $\alpha$ is the polariton-polariton interaction strength; and $R$ is the condensate saturation rate. Results within this paper are not exclusive to this simplified version of the equation of motion and can be reproduced using standard reservoir based models \cite{Wouters_PRL2007}.

We fix the system parameters similar to those in Ref.~[\onlinecite{Keeling_PRL2008}]: $\alpha =2.4$ meV $\mu$m$^2$, $R/\alpha =0.3$, $\Gamma  =0.5$ ps$^{-1}$, and $m = 3 \times 10^{-5} m_0$ where $m_0$ is the free electron mass. All numerical results are performed under the presence of stochastic white noise effectively replicating classical thermal fluctuations. It should be stressed that the multistable regimes and instability points reported are not sensitive to naturally occurring disorder in planar cavities, driving field inhomogeneities~\cite{Sigurdsson_PRB2017}, and energy relaxation due to condensate interactions with the exciton reservoir.
\begin{figure}[t!]
\centering
\includegraphics[width=0.8\linewidth]{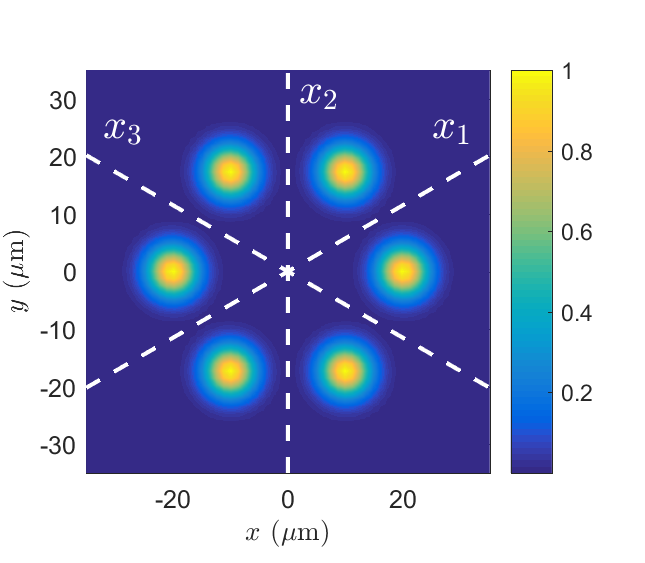}
\caption{Normalized pump profile $P(\mathbf{r})$ of six Gaussian spots arranged into a hexagon with nearest neighbor distance $20$ $\mu$m. Three axis of symmetry between pump spots are plotted with white dashed lines and denoted $x_1$, $x_2$ and $x_3$.}
\label{figHEX0}
\end{figure}
\begin{figure}[t!]
\centering
\includegraphics[width=1\linewidth]{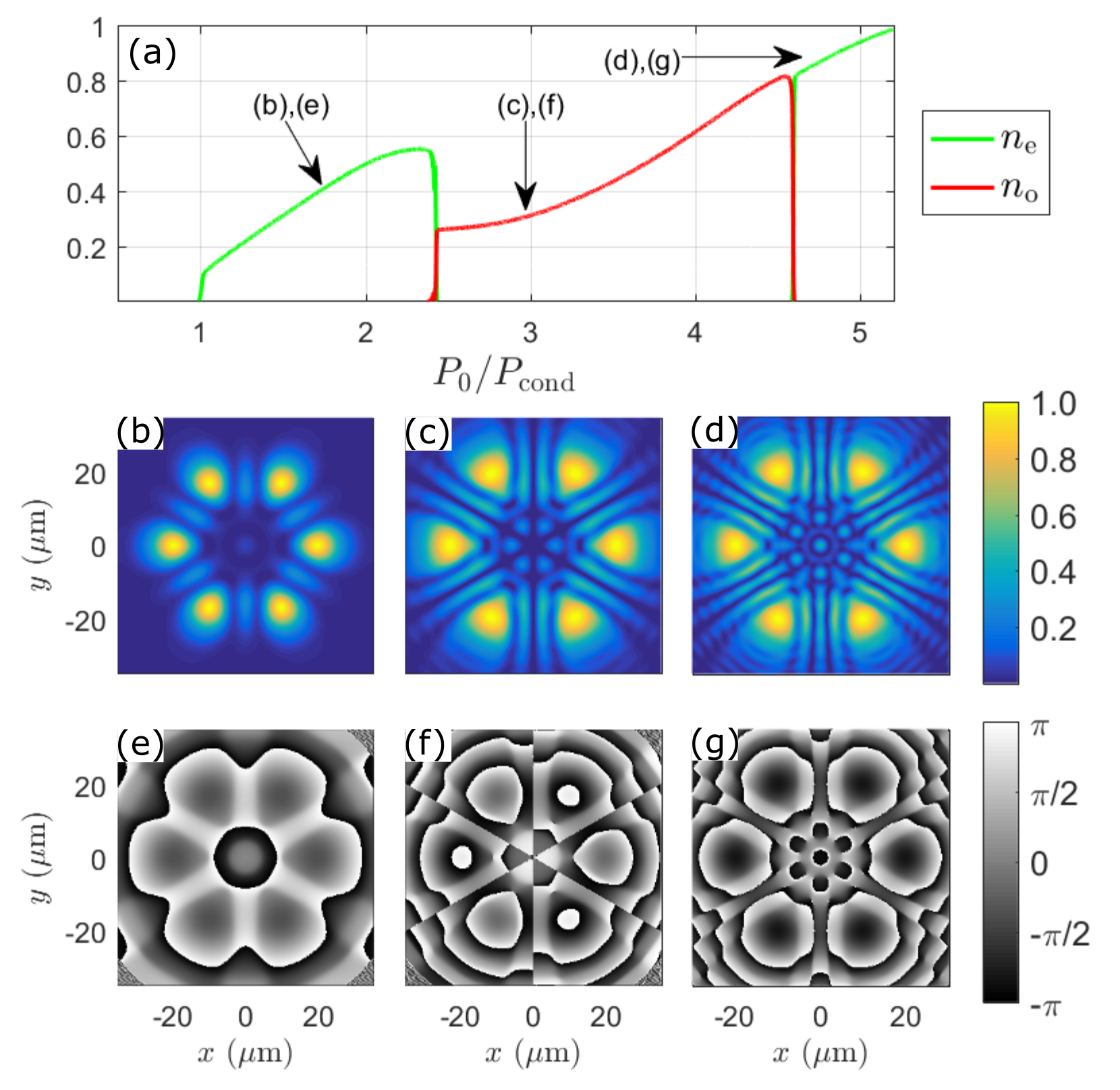}
\caption{(a) Evolution of the normalized parity intensities for a hexagonal pattern of Gaussian pump spots under slow increasing of $P_0$. Sudden transitions between different parity solutions take place at high pump power. Density (b-d) and phase (e-g) plots show the corresponding solution at the points indicated by arrows in panel (a). Here we set $V(\mathbf{r}) = 0$.}
\label{figHEX}
\end{figure}

In order to illustrate the phase locking of different condensates we begin by simulating an open system with $V(\mathbf{r}) = 0$ and $P(\mathbf{r}) \propto P_0$ corresponding to a hexagonal arrangement of Gaussian pump spots with average intensity $P_0$ (see Fig.~\ref{figHEX0}). If the distance between the spots is large enough, no phase locking takes place and each condensate forms with its own phase distribution. When the spots are brought close together, a new condensation threshold of the system belongs to phase locked condensates (see Fig.~\ref{figHEX}). Here we define a phase locking order parameter, or \emph{parity intensity}, as follows,
\begin{align} \label{ne}
n_\text{e} &= \left| \int_A \psi(\mathbf{r}) d\mathbf{r}\right| ^{2}, \\ \label{no}
n_\text{o} &= \left| \int_A \psi(\mathbf{r}) \frac{x_1 x_2 x_3}{|x_1 x_2 x_3|} d\mathbf{r}\right| ^{2}. 
\end{align}
where $x_1$, $x_2$, and $x_3$ are anticlockwise rotated about the $x$-axis by an angle $\pi/6$, $\pi/2$, and $5\pi/6$ respectively (see Fig.~\ref{figHEX0}). Here `e' and `o' stand for {\it even} and {\it odd} parity respectively. As pump intensity is increased the condensates reorganize their phase locking. This type of instability can be regarded as a {\it parity transition} where a condensate with a well defined parity is overcome by a different parity state. High density condensates in such geometric configurations also display vortex-antivortex lattice formation~\cite{Tosi_NatComm2012} which can be seen in the central region of Fig.~\ref{figHEX}g. In what follows we will attempt to shed light on the mechanism of these sudden transitions.

In order to simplify the numerical experiment, and get a clearer understanding on the mechanism causing reorganization of phase locked condensates, we choose an infinite potential quantum well (IQW) as our system.
\begin{equation}
V_0(\mathbf{r}) = \left\{
  \begin{array}{l l}
    0, & \quad \text{if} \ |x|,|y| < L/2, \\
    \infty, & \quad \text{if} \ |x|,|y| \geq L/2. \\
\end{array} \right.
\end{equation}
We set $g = 0$ for simplicity. Let us begin our analysis by choosing a basis satisfying the following real linear part of Eq.~\ref{SE},
\begin{equation} \label{eqH0}
\hat{H}_0 = -\frac{\hbar^2}{2m} \nabla^2 + V(\mathbf{r}).
\end{equation}
Here, separation of variables allows us to write the eigenstates of Eq.~\ref{eqH0} in the form $|\phi_{n_x,n_y} \rangle = | \phi_{n_x} \rangle \otimes | \phi_{n_y} \rangle$ where $|\phi_{n_x}\rangle$ and $|\phi_{n_y}\rangle$ are the eigenstates of the 1D IQW. The solutions in the coordinate basis are written,
\begin{equation} \label{eqEigf}
\phi_{n_x n_y}(\mathbf{r}) = \phi_{n_x}(x) \phi_{n_y}(y),
\end{equation}
where
\begin{align} \label{eqBasis}
\phi_{n_x}\left(x\right) &= \sqrt{\frac{2}{L}} \sin{\left[ n_x \pi \left(\frac{x}{L} - \frac{1}{2} \right) \right]}. \\
\phi_{n_y}\left(y\right) &= \sqrt{\frac{2}{L}} \sin{\left[ n_y\pi \left(\frac{y}{L} - \frac{1}{2} \right) \right]}. 
\end{align}
Similar to the analysis in Refs.~[\onlinecite{sigurdsson2015switching}] and [\onlinecite{Sigurdsson_PRB2017}] we write our time dependent order parameter in the basis of the linear eigenstates:
\begin{equation} \label{eqOP}
\psi(\mathbf{r},t) = \sum_{n_x n_y} c_{n_x n_y}(t) \phi_{n_x n_y}(\mathbf{r}) e^{-i \omega_{n_x n_y} t}.
\end{equation}
Here, $\hbar \omega_{n_x n_y}$ are the linear eigenenergies of Eq.~\ref{eqH0}. For brevity we will write the indexing as $n = (n_x,n_y)$ where $n$ is a unique quantum number for all states. Inserting Eq.~\ref{eqOP} into Eq.~\ref{SE} and integrating out the spatial dependence we arrive at a coupled set of dynamical equations:
\begin{align} \notag
i \hbar \frac{\partial c_{n}}{\partial t} =& \left(\hbar \omega_{n} - i \frac{\hbar \Gamma}{2} \right) c_{n} + i \sum_{m} p_{nm} c_{m} \\ \label{eqCm}
+ & (\alpha - i R) \sum_{n jkl} M_{njkl} c_j^* c_k c_l.
\end{align}
Here,
\begin{equation} \label{eqPu}
p_{nm} = \int_A P(\mathbf{r}) \phi_{n_x n_y}^* \phi_{m_x m_y} d \mathbf{r},
\end{equation}
and
\begin{equation}
M_{njkl} = \int_A  \phi_{n_x n_y}^* \phi_{j_x j_y}^* \phi_{k_x k_y} \phi_{l_x l_y}d \mathbf{r}.
\end{equation}
It becomes immediately clear that if $P(-x,y) = P(x,y)$ and $P(x,-y) = P(x,y)$ then only states $\phi_{n_x,n_y}$ of the same parity structure along $x$ and $y$ coordinate are coupled by the pump gain mechanism (Eq.~\ref{eqPu}). As an example, the state $\phi_{22}$ does not contribute to the gain of $\phi_{11}$ but does however affect its decay through the nonlinear saturation $R$. It then becomes evident that the condensate threshold belongs to a superposition of same parity eigenstates $c_n$. Furthermore, it is clear that degenerate states such as $\phi_{12}$ and $\phi_{21}$ evolve equivalently.

The last term of Eq.~\ref{eqCm} is analogous to polarization coupling of electromagnetic waves in media with $\chi^{(3)}$ Kerr nonlinearity. The elements $M_{njkl}$ only depend on the choice of basis to represent the condensate order parameter. Evidently, since $\phi_{n_x n_y}$ are representable as real functions, the order of the indices in $M_{njkl}$ does not matter but it does matter for the evolution of the amplitudes $c_m$ (Eq.~\ref{eqCm}) which are coupled through the sum of products $M_{njkl} c_j^* c_k c_l$, mixing the phases of different parities. It becomes clear that without nonlinearity the condensate cannot be stable above threshold as it will diverge, but it is also clear the the nonlinearity term is the only one responsible to cause instability points shown in Fig.~\ref{figHEX}a where one condensate solution is replaced by another. More interestingly, only the imaginary nonlinear term ($R$) is enough to produce these unstable points~\cite{Sigurdsson_PRB2017}.

The nonlinear problem is therefore highly nontrivial and searching for multistable solutions relies heavily on numerical modeling. However, a qualitative result on pump induced, parity dependent, instability was given in Ref.~[\onlinecite{Sigurdsson_PRB2017}] where the critical instability point was demonstrated analytically for a two-mode system. The full quenching (parity transition), as shown in Fig.~\ref{figHEX}a, where the green colored curve is replaced by red and vice versa, is then made possible when the critical point gives rise to a solution capable of driving the previous solution to zero, a feature made possible through the nonlinear saturation. The reorganization of phase locked condensates shown in Fig.~\ref{figHEX} can therefore be regarded as parity transitions within a single spatially modulated condensate.

Furthermore, since the gain of the condensate depends on the condensed mode in question through the saturation term $R|\psi|^2$, it has been shown that a hysteresis can be recovered when the pump intensity is swept below the point of instability, evidencing \emph{parity bistability}~\cite{sigurdsson2015switching}. Here, in contrast to bistable 1D states, the degeneracy of the 2D linear equations of motion results in multistability.

\section{Results}
We continue our analysis in the IQW system. We work with the following pump profile:
\begin {equation} \label{eq1}
P(\mathbf{r})=P_0 \left| \phi_{22} (\mathbf{r})\right| ^{2}.
\end {equation}
Here, $P_0$ represents the average nonresonant driving field intensity. Similar to Eqs.~\ref{ne} and \ref{no} we define new parity intensities for each possible parity structure along the $x$- and $y$-axis of the system:
\begin{equation}\label{parity}
\begin{aligned}
n_\text{ee} &= \left| \int_A \psi(\mathbf{r}) d\mathbf{r}\right| ^{2}, \\
n_\text{eo} &= \left| \int_A \psi(\mathbf{r})\frac{y}{|y|} d\mathbf{r}\right| ^{2}, \\
n_\text{oe} &= \left| \int_A \psi(\mathbf{r}) \frac{x}{|x|}d\mathbf{r}\right| ^{2}, \\
n_\text{oo} &= \left| \int_A \psi(\mathbf{r}) \frac{xy}{|x y|} d\mathbf{r}\right| ^{2}.
\end{aligned}
\end{equation}
When $\psi$ is in a superposition of modes of only one parity structure then only one of the above integrals is non-zero. We note that the choice of driving field $P(\mathbf{r})$ is non-exhaustive and Eq.~\ref{eq1} is only one example of many to create a spatially patterned condensate and produce similar results to this paper.

\subsection{Multistability}
\begin{figure}[t!]
\centering
\includegraphics[width=\linewidth]{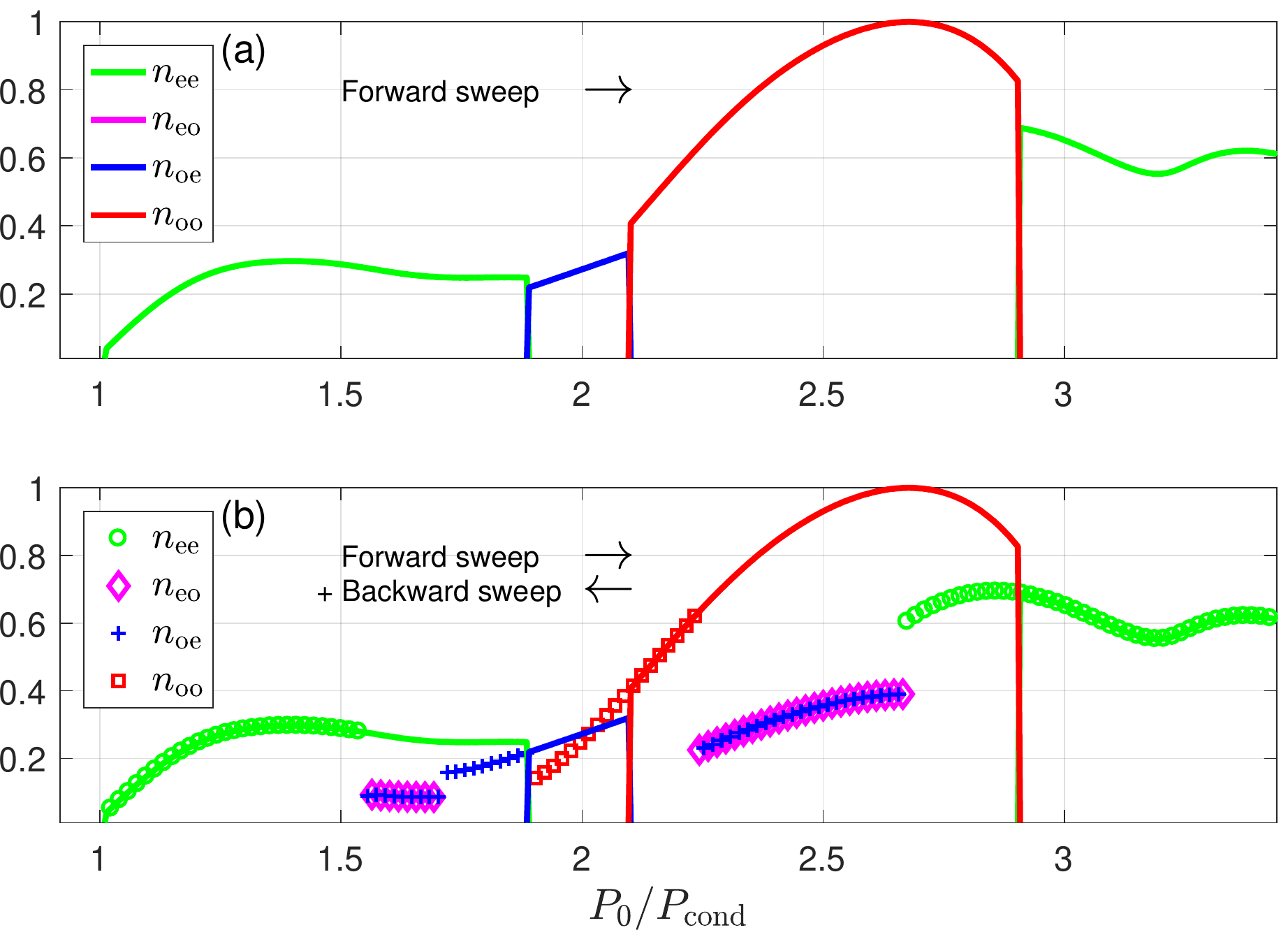}
\caption{(a) Evolution of the normalized parity intensities when $P_0$ is slowly swept to higher values. Points of instability (parity transitions) take place when one colored curve is replaced by another. (b) Forward (whole lines) and backwards (markers) sweep of $P_0$ give away hysteresis regions indicating the bistable nature of the condensate.}
\label{fig1}
\end {figure}

\begin{figure}[t!]
\centering
\includegraphics[width=1\linewidth]{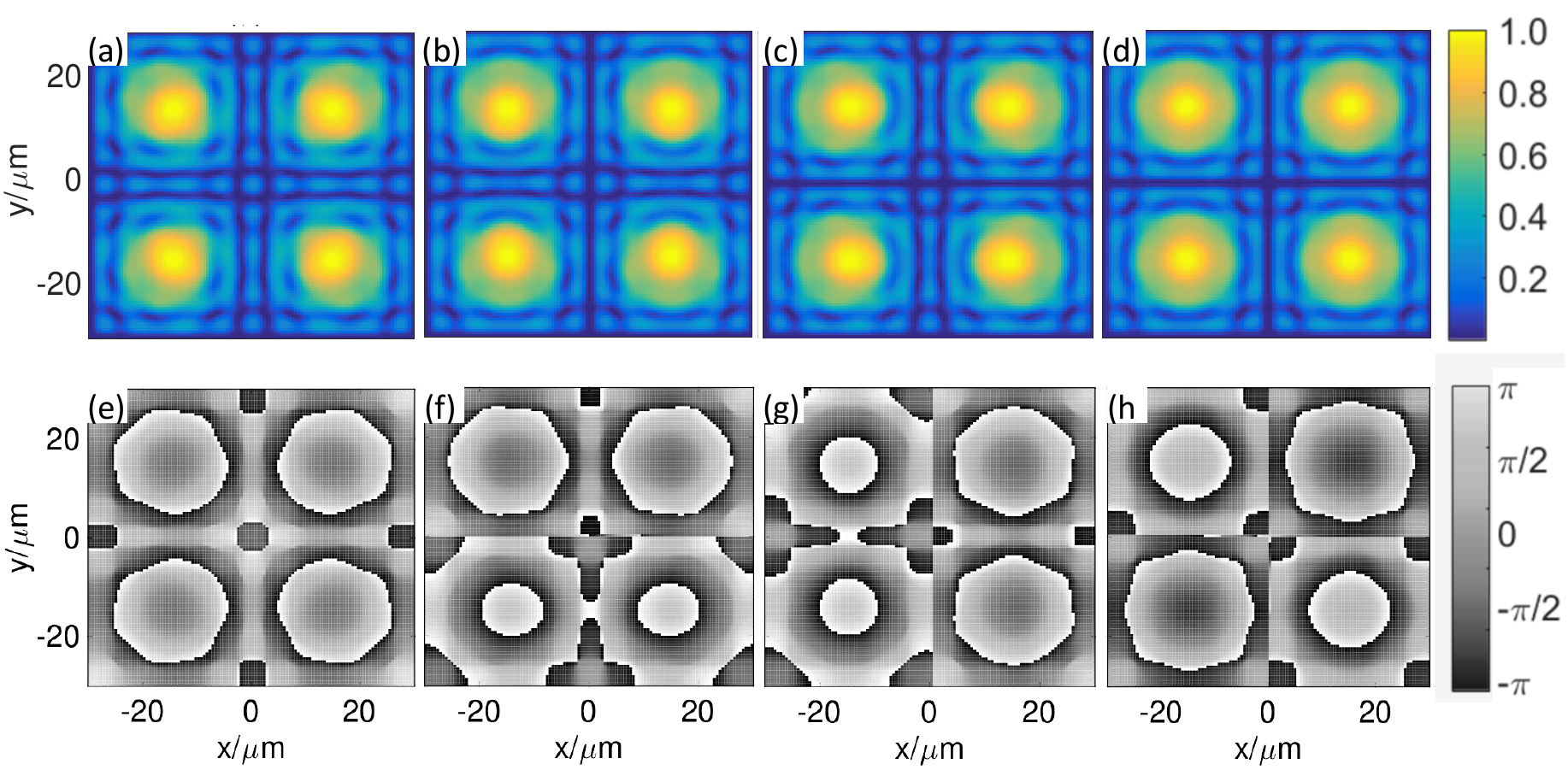}
\caption{Four steady states within the square trap using Eq.~\ref{eq1} at the same excitation power $P_0/P_\text{cond} =2.75$: (a-d) Normalized condensate density and (e-h) phase corresponding to $n_\text{ee}$, $n_\text{eo}$, $n_\text{oe}$ and $n_\text{oo}$ steady states respectively.}
\label{fig2}
\end {figure}

In Fig.~\ref{fig1}a we show the results of sweeping adiabatically forward in pump intensity. As expected, we observe the clear phase locking upon condensation corresponding to $n_{ee}$ (symmetric state) as the initial dominant parity structure. Increasing the pump intensity further we observe recurrent drops in the parity intensities of the condensate quickly replaced by another solution of different parity, corresponding to parity cross-saturation analogous to the observed drops in Fig.~\ref{figHEX}. In Fig.~\ref{fig1}b we sweep backwards in pump intensity (colored markers) and find several hysteresis intervals indicating the bistable nature of the system. As an example, at $P_{0}/P_\text{cond} = 2.75$, the system can exist in at least two steady states, separately characterized by having $n_\text{ee}$ (green circles) and $n_\text{oo}$ (red whole curve) as the dominant parity intensities. These states are analogous to the observations of Ref.~[\onlinecite{Tosi_NatComm2012}] except without vortex lattices (see Fig.~\ref{fig2}[a,d]), classifying them as ferromagnetic (FM) and antiferromagnetic (AFM) ordered states in phase. However, two degenerate anisotropic states can also exist which have not been reported before, to our knowledge, and correspond to the blue and magenta coloring in Fig.~\ref{fig1} (also Fig.~\ref{fig2}[b,c]). It is clear that these states are interchangeable since the $n_\text{eo}$ and $n_\text{oe}$ parity structures are one and the same under $\pi/2$ rotation of the system. Due to this degeneracy the condensate can form a solution as a superposition of $n_\text{eo}$ and $n_\text{oe}$ states as shown in Fig.~\ref{fig1}b where blue and magenta markers co-exist.

Sweeping forward and backward in pump intensity reveals bistable intervals but does not elucidate the multistability of the system. By the virtue of symmetry between $x$ and $y$ it is clear that if $n_\text{ee}$ and $n_\text{oo}$ can exist at the same pump power, then $n_\text{eo}$ and $n_\text{oe}$ are also possible stable states. This is verified numerically by activating $P_0$ instantaneously and rapidly condensing the system from a stochastic initial condition, resulting in one of the four different possible stable states (see Fig.~\ref{fig2}). We note that due to the spatial structure of $P(\mathbf{r})$ no vortices become present within the system as opposed to the case of uniform pump profiles~\cite{Borgh_PRB2012, Liew_PRB2015}. Density and phase plots of the four different multistable states are shown in the upper and lower rows of panels in Fig.~\ref{fig2}, respectively.

\subsection{Parity bifurcation}
\begin{figure}[t!]
\centering
\includegraphics[width=0.8\linewidth]{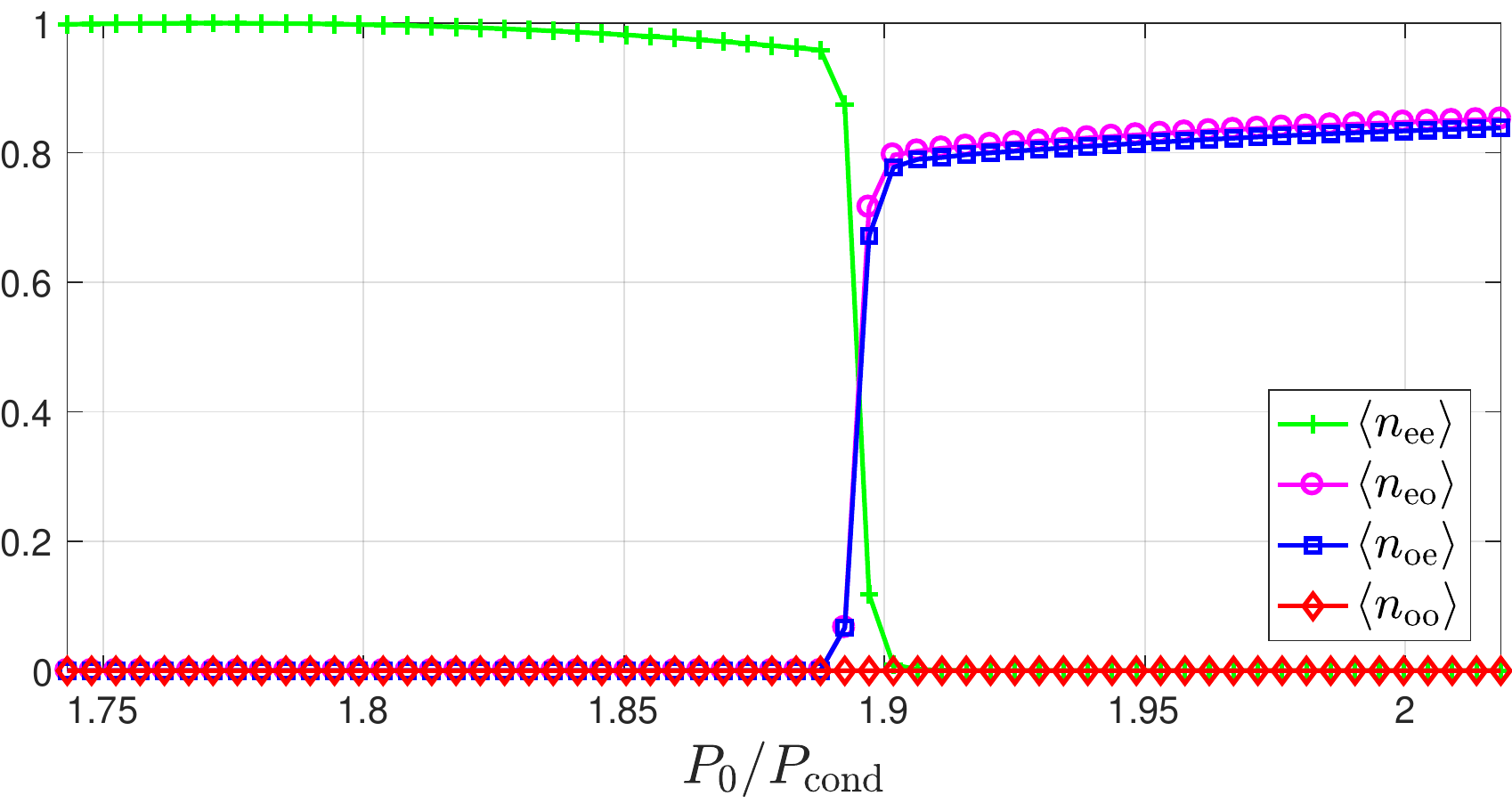}
\caption{Average normalized parity intensity of the trapped condensate over 1000 stochastic trials. As the $P_0$ is adiabatically increased the condensate bifurcates into one of two possibilities corresponding to states $n_\text{eo}$ and $n_\text{oe}$.}
\label{figBif}
\end{figure}

Earlier it was stated that the $n_\text{oe}$ and $n_\text{eo}$ solutions evolve in an indistinguishable manner due to the symmetry of the system. We verify this by averaging over 1000 stochastic trials (Monte Carlo) simulating the evolution of the condensate as a function of pump power up to the point of instability (bifurcation point) where the condensate transitions to either an $n_\text{eo}$ or $n_\text{oe}$ state. The trials are stochastic by introducing a white noise field to the order parameter at much smaller time steps than the polariton lifetime, keeping any uncondensed states completely stochastic in evolution. In Fig.~\ref{figBif} the average parity intensity drops from the $n_\text{ee}$ state and populates either an $n_\text{eo}$ or $n_\text{oe}$ state with approximately equal probability. Such process can be classified as \emph{parity bifurcation}.

We stress that the critical transition in Fig.~\ref{figBif} depends on the adiabaticity of the pump. It is clear that if the $n_\text{eo}$ and $n_\text{oe}$ solutions become unstable and condense then $n_\text{oo}$ should also be unstable. Stochastic fluctuations can not only seed the odd parity along one of the system axes, resulting in an $n_\text{eo}$ or $n_\text{oe}$ state, but also seed odd parity states along both the $x$ and $y$ axes close enough in time that a $n_\text{oo}$ state is recovered. Such a process is however only likely if the adiabaticity of the pump is relaxed (faster ramping rates). The parity bifurcation naturally vanishes when disorder is introduced, corresponding to breaking of the $\pi/2$ invariance of the system, making one of the two states deterministically dominant. We stress that the influence of disorder only breaks the equivalence of $n_\text{eo}$ and $n_\text{oe}$ states without affecting multistable regimes between $n_\text{ee}$, $n_\text{eo,oe}$, and $n_\text{oo}$ states.

\subsection{\label{transition22} Controlled transitions between multistable states}
A sweep of the pump intensity can be utilized to switch between different states as is evident in Fig.~\ref{fig1} but a more pragmatic method relies on controllably perturbing the system to deterministically change between states. By applying short nonresonant pulses to create pressure gradients within the condensate, controlled transitions between the four steady states can be induced. We will demonstrate that the transitions (or {\it steps} for brevity) based on these nonresonant pulses can be made from any of the four steady states to any of the other steady states. Therefore any path made up of these steps is completely reversible.

We present here three possible steps (not counting backwards steps), namely $n_\text{ee} \leftrightarrow n_\text{eo}$, $n_\text{ee} \leftrightarrow n_\text{oe}$, and $n_\text{ee} \leftrightarrow n_\text{oo}$. With this set of steps, it is possible to transit between all of the four states with at most two steps. The pump pulse is simulated by replacing the $iP\left( \mathbf{r} \right)$ term with $i\left[ P\left( \mathbf{r} \right) +P'\left ( \mathbf{r},t \right) \right]$ where $P'$ is a short pulse superimposed upon the static excitation,
\begin {equation}
\label {pulse}
P'\left( \mathbf{r},t\right) = P'_{0} e^{-\left( \frac{t-t_{0}}{\delta t} \right) ^{2}} p(\mathbf{r})_{(A,B,C,D)},
\end {equation}
where $P'_{0}$ determines the intensity of the pulse and $\delta t=20$ ps determines the temporal length of the pulse. For simplicity, we choose the following scheme of pulsing:
\begin{align} \notag
p(\mathbf{r})_{(ABCD)} & =  A \frac{(1-x/|x|)(1+y/|y|)}{4}  \\ \notag
&+  B \frac{(1 + x/|x|)(1+y/|y|)}{4}  \\ \notag
&+  C \frac{(-1+x/|x|)(-1+y/|y|)}{4}  \\
&+  D \frac{(1+x/|x|)(1-y/|y|)}{4}. \label{eqp}
\end{align}
Here $x,y \in [-L/2,L/2]$ and $A,B,C,D \in [0,1]$. The function $P'(\mathbf{r},t)$ corresponds then to a short duration of added uniform gain in the chosen quadrants of the quantum well. The choice of which quadrants are to be excited is given by the binary control variables $A$, $B$, $C$ and $D$ corresponding to a pulse being activated in the top right, top left, bottom left and bottom right quadrants of the square well respectively. We stress that the number of methods of perturbing the system is nonexhaustive and the one given by Eq.~\ref{eqp} is just one of many possibilities.
\begin{figure}[t!]
\centering
\includegraphics[width=\linewidth]{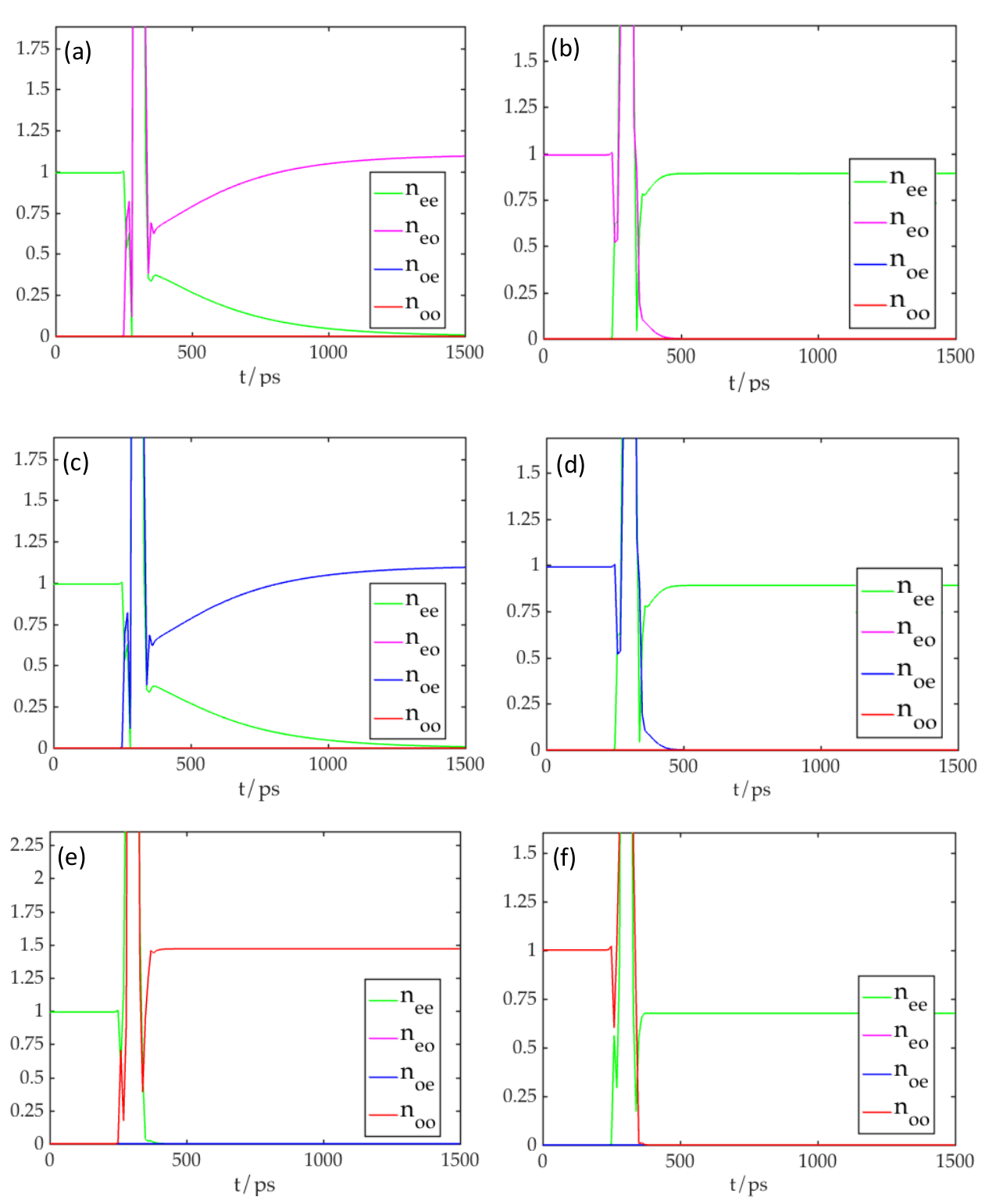}
\caption{Time evolution of the normalized parity intensities of the polariton condensate being perturbed at $t = 300$ ps with a (a,b) $p_{(1001)}$, (c,d) $p_{(1100)}$, and (e,f) $p_{(1010)}$ pulse, resulting in the condensate abandoning its old parity structure and settling into a new one. (a) $n_\text{ee} \rightarrow n_\text{eo}$, (b) $n_\text{eo} \rightarrow n_\text{ee}$, (c) $n_\text{ee} \rightarrow n_\text{oe}$, (d) $n_\text{oe} \rightarrow n_\text{ee}$, (e) $n_\text{ee} \rightarrow n_\text{oo}$, (f) $n_\text{oo} \rightarrow n_\text{ee}$.}
\label{pulse}
\end{figure}

Three pulse profiles are employed, $p_{(1001)}$, $p_{(1100)}$, and $p_{(1010)}$ for the $n_\text{ee} \leftrightarrow n_\text{eo}$, $n_\text{ee} \leftrightarrow n_\text{oe}$, and $n_\text{ee} \leftrightarrow n_\text{oo}$ transitions respectively. The results are presented in Fig.~\ref{pulse}. Let us analyze one transition as an example. The steps $n_\text{eo} \rightarrow n_\text{ee}$ and $n_\text{eo} \rightarrow n_\text{ee}$ (see Fig.~\ref{pulse}[a,b]) are both induced using a $p_{(1001)}$ profile. This is a logical choice of pump pulse profile since the condensate is being perturbed in the $x>0$ quadrants, evolving differently from the rest of the condensate at $x<0$. The perturbing dynamics are thus symmetric about the $x$-axis but encourage asymmetry about the $y$-axis. In Fig.~\ref{pulse}a we see that the $n_\text{ee}$ state decays much slower as opposed to the decay of the $n_\text{eo}$ state in Fig.~\ref{pulse}b indicating that the transition dynamics are distinguishable. Numerical simulations over long times reveal that the $n_\text{ee}$ intensity eventually decays down to the white noise level of the simulation.

The other two transitions pairs ($n_\text{ee} \leftrightarrow n_\text{oe}$ and $n_\text{ee} \leftrightarrow n_\text{oo}$) can be induced in a similar fashion with the $p_{(1100)}$ and $p_{(1010)}$ pump profiles respectively. Our system therefore possesses a cyclical transition scheme that allows for controlled transitions between all four steady states.

\subsection{Effects of pump induced blueshift}
\begin{figure}[t!]
\centering
\includegraphics[width=1\linewidth]{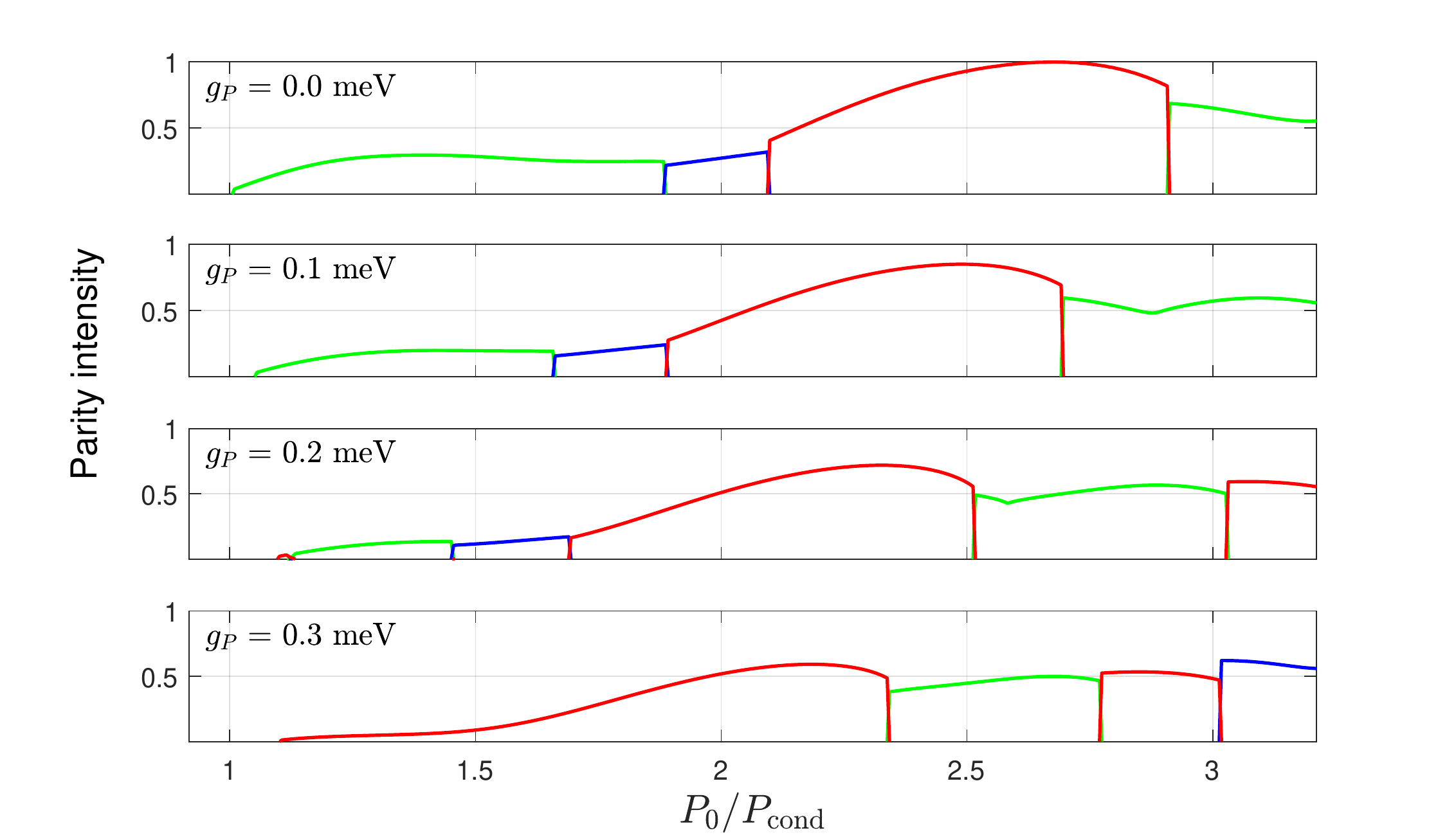}
\caption{Forward sweep of $P_0$ for different values of $g$. Green, blue, and red curves correspond to $n_\text{ee}$, $n_\text{eo,oe}$, and $n_\text{oo}$ parity intensities respectively. The value $g_P$ corresponds to maximum blueshift at $P_0 = 2 P_\text{cond}$ where $P_\text{cond}$ is defined as the condensation value for $g=0$.}
\label{figBlue}
\end{figure}

We shift our attention to the case when $g \neq 0$. The real potential $gP(\mathbf{r})$ alters the energy landscape of the system but is determined by the symmetry of the pump, so, as before, only same parity structures are coupled together in both energy and gain coming from the pump (see Eq.~\ref{eqPu}). Mixing between different parity structures is therefore still only realized through the system nonlinearity which is unchanged. Previous results are thus only quantitatively changed with all parity related effects still present. This is demonstrated in Fig.~\ref{figBlue} where for realistic values of pump blueshift the points of parity transitions are shifted along the $P_0$ axis. We define the blueshift parameter as $g_P = \text{max}{\left[ g P(\mathbf{r}) \right]}$ for $P_0 = 2P_\text{cond}$, that is, the maximum energy shift at twice threshold intensity.

\subsection{Generalization to symmetric pump and trapping geometries}
\begin{figure}[t!]
\centering
\includegraphics[width=1\linewidth]{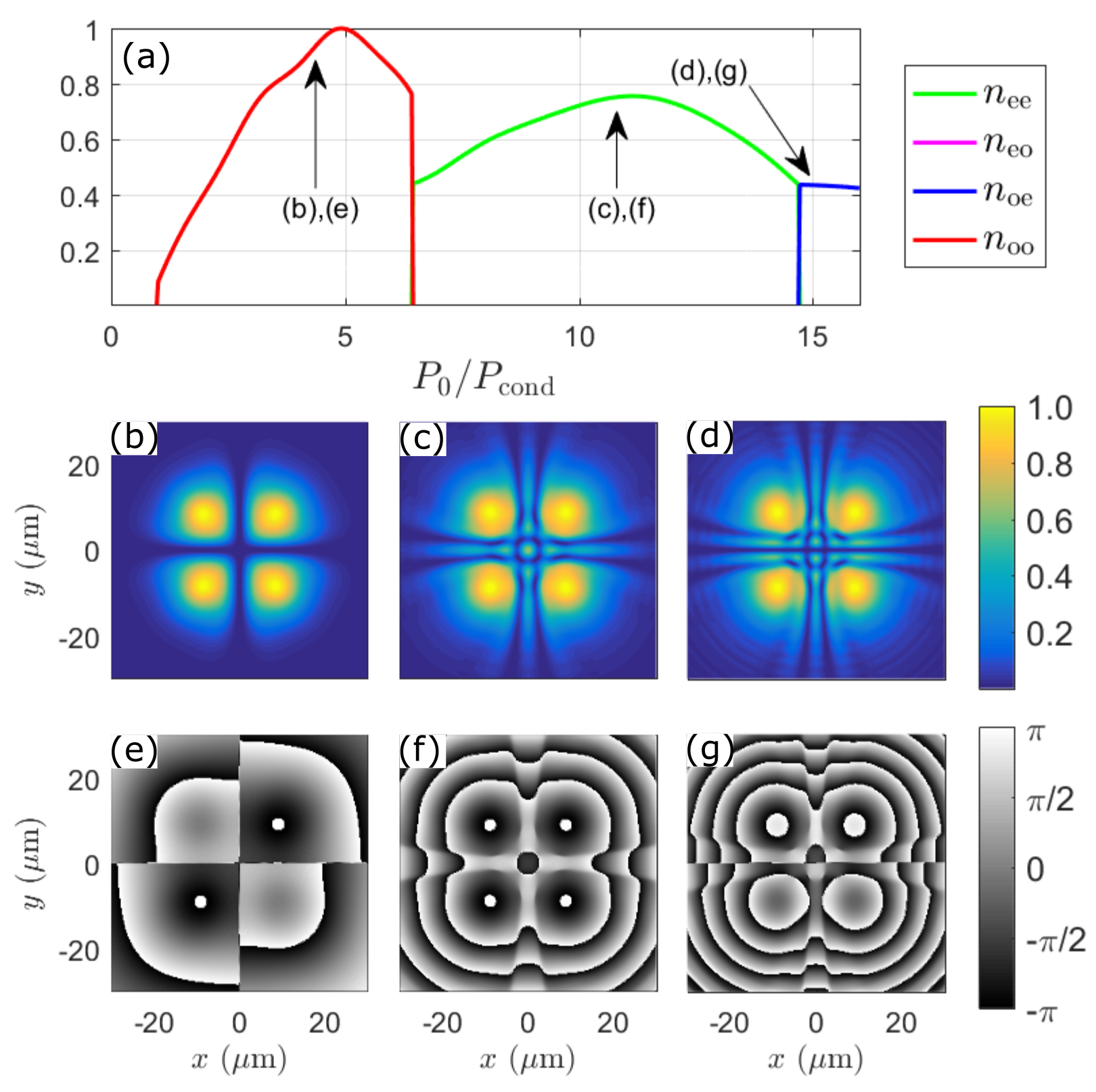}
\caption{(a) Evolution of the normalized parity intensities for the harmonic potential, $V(\mathbf{r}) = u (x^2 + y^2)$, where $u = 0.2$ $\mu$eV, and $P(\mathbf{r}) = P_0 |\phi_{22}(\mathbf{r})|^2$, showing also clear transitions between the parities of the condensate (colored lines). Normalized density (b-d) and phase (e-g) plots show the corresponding solution at the points indicated by arrows in panel (a).}
\label{figHO}
\end{figure}

The above results are naturally not exclusive to the IQW and can mostly be replicated using other symmetric trapping geometries $V(\mathbf{r})$ or driving-field profiles $P(\mathbf{r})$ such as the hexagonal pattern shown in Fig.~\ref{figHEX0}. In fact, any real potential and driving field in the planar system which are symmetric about the $xy$ coordinates will never mix together the different parity structures. As an example, the harmonic oscillator is a cylindrically symmetric potential and does not distinguish the $x$ and $y$ coordinates. When the system is driven by a symmetric nonresonant field $P(\mathbf{r}) = P_0 |\phi_{22}(\mathbf{r})|^2$ parity transitions analogous to Fig.~\ref{fig1} are observed in the polariton condensate (see Fig.~\ref{figHO}).

\begin{figure}[t!]
\centering
\includegraphics[width=1\linewidth]{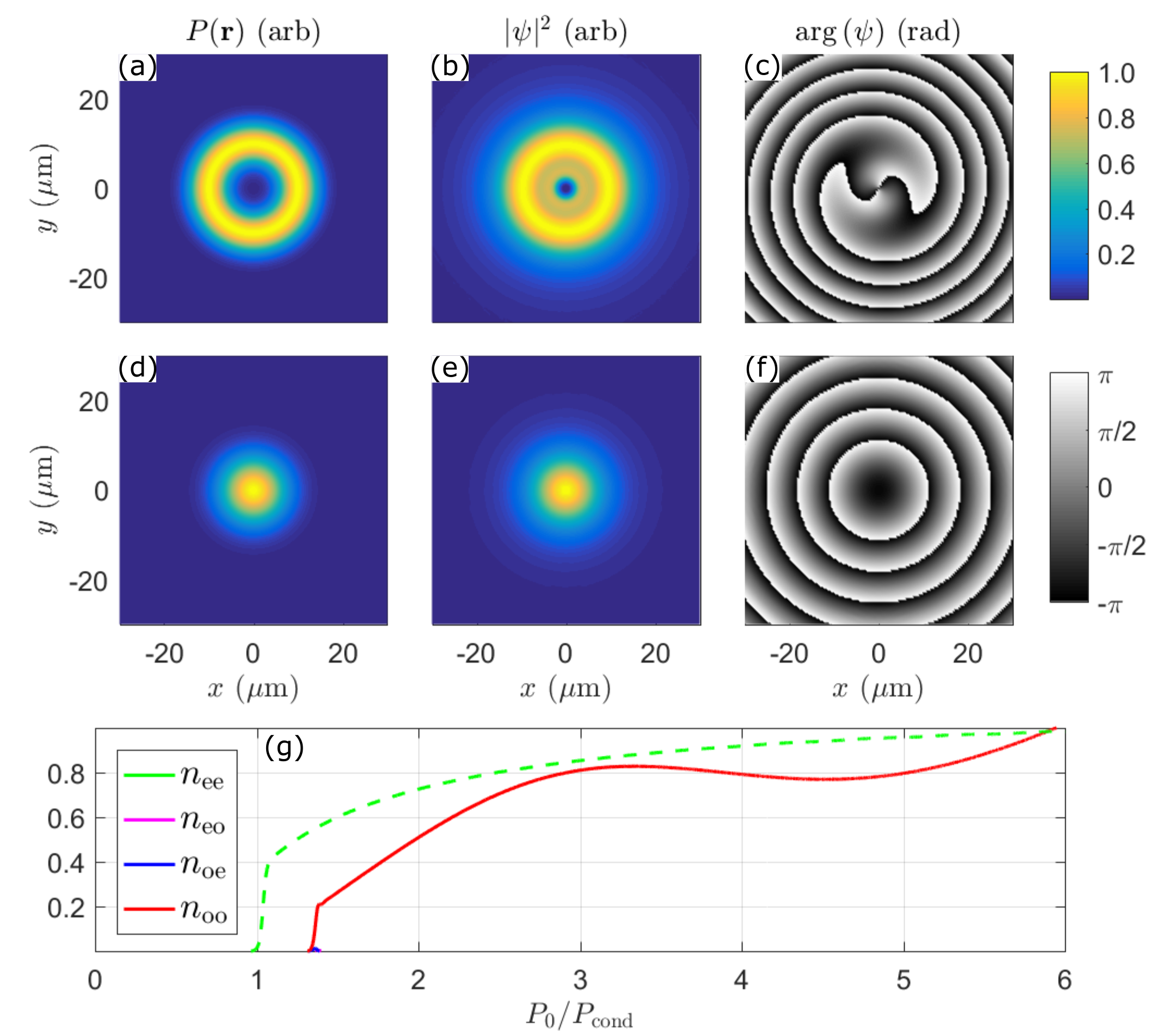}
\caption{(a) Normalized annular shaped pump profile resulting in a (b) circular density (normalized to same colorscale) profile of the polariton condensate with a circular flow (c) corresponding to a vortex of topological charge $|m| = 2$. (d) Gaussian shaped pump profile resulting in a (e) Gaussian shaped polariton condensate (f) with only radial flow of polaritons. (g) Normalized parity intensities for annular pump (solid curve) and Gaussian pump (dashed curve) across $x$ and $y$ system axis show no sharp changes in contrast with previous results. Here $V(\mathbf{r}) = 0$ in all plots.}
\label{figAn}
\end{figure}

We also demonstrate certain excitation schemes which do not show the sudden change in parities of the system. It is clear that the competition between the parities to dominate the condensate ultimately depends on how the real and complex potential couple together same parity modes, and subsequently how the modes interact through the nonlinearity of the condensate pattern. Clearly some potentials will favor this conflict between the parities as shown above, whereas others show no sudden transitions in the condensate parity. As an example, annular pump shapes are used to create cylindrically symmetric condensates~\cite{Askitopoulos_PRB2013} to investigate the onset of spontaneous currents~\cite{Liu_PNAS2015}, vorticity~\cite{Li_PRA2016}, and petal formation~\cite{Manni_PRL2011, Dreismann_PNAS2014, Sun_ArXiv2016}. Here, a single annular shaped pump results in a single condensate with well defined phase corresponding to circulating current (see Fig.~\ref{figAn}[a-c]). In the absence of other condensates no phase locking is possible and no instability appears from interference.

The same goes for single Gaussian shaped pump spots resulting in a single condensate with nothing to phase lock with (see Fig.~\ref{figAn}[d-f]). The transition between parities is therefore only observable by appropriate choice of $V(\mathbf{r})$ and $P(\mathbf{r})$, creating a condensate pattern (or likewise, pattern of condensates) which allows phase locking.

\begin{figure}[t!]
\centering
\includegraphics[width=1\linewidth]{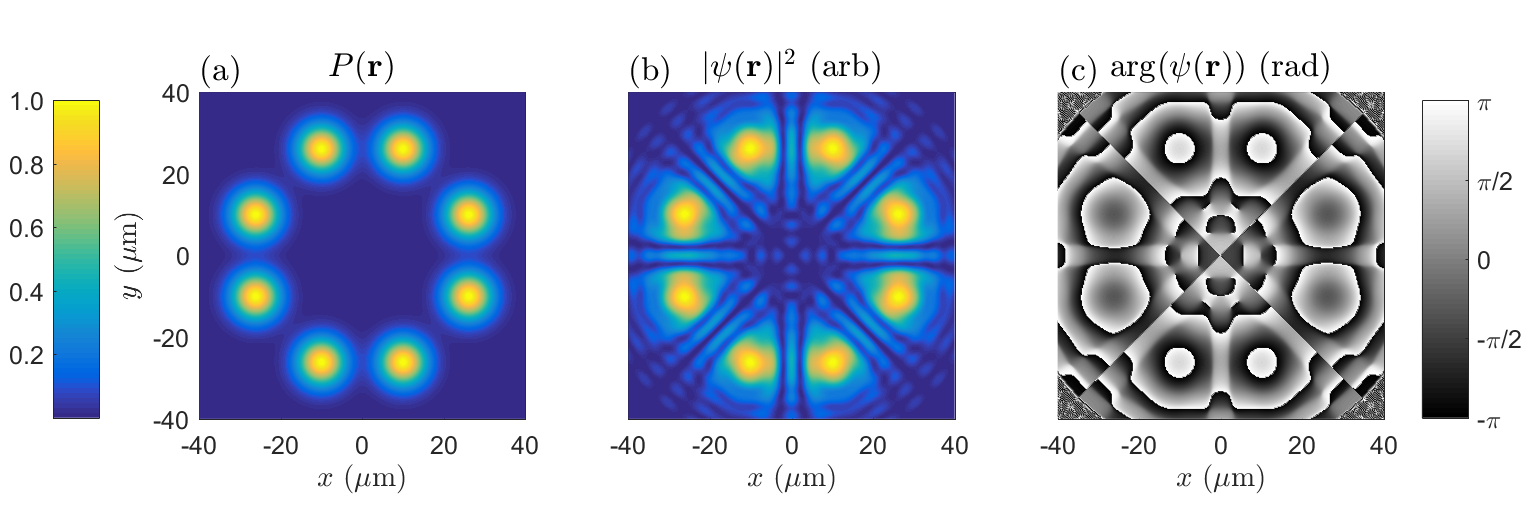}
\caption{(a) Normalized driving field profile of 8 Gaussians arranged in an octagon. (b) Resulting normalized condensate density and (c) phase at a fixed pump intensity showing a steady state of phase locked condensates forming symmetric and antisymmetric bonds interchangeably around the octagon.}
\label{figOCT}
\end{figure}

Lastly, we come back to Fig.~\ref{figHEX} of the hexagonal driving field pattern and address the observation that only two parity structures are visible, $n_\text{e}$ and $n_\text{o}$, which correspond to all nearest neighbors forming a symmetric state or antisymmetric state respectively, and why no other parity arrangements are observed. It is clear that the above analysis is not fit for two of the three axes of symmetry $x_1$ and $x_3$ as shown in Fig.~\ref{figHEX0}. These axes have projections on the $x$ and $y$ coordinates which makes the hexagonal parity structure non-decomposable onto the Cartesian basis. But this is not the reason why only $n_\text{e}$ and $n_\text{o}$ are observed. In fact, using an octagonal pattern of Gaussians results in a steady state with nearest neighbors forming symmetric and antisymmetric bonds along the lattice (see Fig.~\ref{figOCT}), a state not observed when using hexagonal lattices. 

The reason for this observation stems from the fact that the energy and gain of each condensate depends on the number of nearest neighbor symmetric and antisymmetric bonds which, in the hexagonal geometry, can only form a steady state with all bonds being antisymmetric or symmetric to avoid frustration. For example, in Fig.~\ref{fig2} every solution has the same configuration of symmetric and antisymmetric bonds in each quadrant of the system, making them energetically equivalent. The symmetric and antisymmetric bonds can be regarded as a type of FM ($0$) and AFM ($\pi$) bond ordering amongst scalar ~\cite{Tosi_NatComm2012} and spinor polariton condensates~\cite{Ohadi_PRL2017}. Fig.~\ref{figOCT} then shows an example of a lattice favoring a periodic arrangement of FM and AFM bonds purely due to nonlinear interactions since the system is completely isotropic. We point out that, as expected, a lattice of vortices forms within the center of the octagon just like in the hexagonal geometry.

\section{Conclusions}
We have shown through extensive numerical simulation of the complex nonlinear Schrödinger equation that multistability in planar microcavity exciton-polariton condensates, with spatially patterned nonresonant pumping, can be achieved between symmetric and antisymmetric solutions of the condensate pattern. Solutions of different parity structures compete for the system gain with ultimately one winner driving all other solutions to zero through the condensate saturation mechanism. Most notably, doubly degenerate condensate solutions give rise to bifurcating points where the condensate transitions to one of either degenerate states with equal probability at a critical pump power. The bulk of the results focus on $\pi/2$ symmetric linear equations of motion, such as the planar infinite quantum well, where the linear physics of the system is decomposable onto the orthogonal Cartesian basis. But systems with axes of symmetry at angles $<\pi/2$ such as hexagons and octagons also display the same competition and quenching between symmetric and antisymmetric bonded states.

Within the multistable regime, the use of controlled nonresonant pulses allows deterministic switching between different parity structures of the phase locked condensate pattern. Applying the same pulse repeatedly causes the system to alternate between two states. For such pairs of states, the pulse acts effectively as a NOT gate, paving the way towards controllable optical quantum fluid circuitry.

Consequently, a possible future direction would be to consider multiple trapped condensates and attempt to engineer an AND gate. Together, the two gates form a universal set of logic gates for binary processing. Although this relies on a binary logic, a further extension can be to attempt to create a universal set of quaternary logic gates from multiple trapped/localized condensates.

\section*{Acknowledgements}
H.S. acknowledges support by the Research Fund of the University of Iceland, The Icelandic Research Fund, Grant No. 163082-051. T.L. acknowledges support from the Ministry of Education (Singapore) grant 2015-T2-1-055.

\bibliography{Bibliography}

\end{document}